# Geographical Distribution of Biomedical Research in the USA and China

Yingjun Guan
School of Information Sciences
University of Illinois at Urbana-Champaign
Champaign, IL 61820, USA
yingjun2@illinois.edu

Jing Du
School of Information Sciences
University of Illinois at Urbana-Champaign
Champaign, IL 61820, USA
jingdu4@illinois.edu

Vetle I. Torvik
School of Information Sciences
University of Illinois at Urbana-Champaign
Champaign, IL 61820, USA
vtorvik@illinois.edu

## ABSTRACT

We analyze nearly 20 million geocoded PubMed articles with author affiliations. Using K-means clustering for the lower 48 US states and mainland China, we find that the average published paper is within a relatively short distance of a few centroids. These centroids have shifted very little over the past 30 years, and the distribution of distances to these centroids has not changed much either. The overall country centroids have gradually shifted south (about 0.2° for the USA and 1.7° for China), while the longitude has not moved significantly. These findings indicate that there are few large scientific hubs in the USA and China and the typical investigator is within geographical reach of one such hub. This sets the stage to study centralization of biomedical research at national and regional levels across the globe, and over time.

• Applied computing → Digital libraries and archives

## 1 INTRODUCTION

During the past few decades, there has been an explosive growth and geographical spread of the scientific literature [1]. To explore these changes and create an updated framework for innovation-oriented subjects and federal funds, there has been an intense research activity in studying the geographic distribution of scientific activities [2]. A variety of studies have examined city concentrations [3,4], countries [5,6,7], subject areas or journals [8,9], and innovation as reflected in patents [10,11,12] and industrial activity [13]. This research suggests that linkages between research affiliations are fostered by geographic proximity, and geographic distance is an obstructive factor in achieving collaborations. Therefore, it is of significance to analyze the geographical proximity of localities to regional hubs.

To analyze the geographical proximity and movement of biomedical research in the USA and China, this study focuses on geocoded author affiliations of PubMed articles [14] published in the past 30 years by authors from the lower 48 states of the USA and mainland China. Both countries are top producers of articles, and constitute large geographical areas and geographical diversity, which make them interesting subjects of study here. The research hubs for each country, as represented by centroid longitude-latitude pairs, are identified at several different levels of granularity using K-means clustering of Vincenty distances. The centroid proximity distributions are examined to characterize movement and geographical proximity relative to centroid representations of these research hubs.

## 2 DATA

The most recent version of MapAffil [14] contains disambiguated and geocoded place names of 37 million affiliations from nearly 20 million PubMed articles published between 1867 and 2016. It covers a significant portion affiliations missing from PubMed. These were harvested from external sources including PubMed Central, Microsoft Academic Graph (MAG), Astrophysics Data System (ADS), and NIH grants. The geographical data of each article is identified by MapAffil, which maps an author's affiliation to its city and the corresponding city-center geocode (the longitude and latitude) across 227 countries and territories worldwide. MapAffil has a high overall performance and provides additional geo-linked data e.g., via US FIPS codes. It is worth noting that place name disambiguation is a non-trivial process complicated by a) many relatively rare place names with low prevision such as the ones listed in Table 1, and b) the ambiguity of highly frequent place names, such as the ones listed in Table 2. The precision and recall measures listed in Table 2 reflect the performance of identifying all instances of the actual city when using a blatantly naïve approach: exact match on its (case-insensitive) place name. Place names have low precision (when the name points to multiple places or other things that are not place names; "New York" can refer to the US state or the city). Conversely, place names have low recall (when the place has variant names or sub-divisions such as the boroughs of New York City). Table 2 also shows that the top cities in each country produce at least twice as many papers as their population proportion and many cities outside the US produce of their country's output (Paris, France: 39.5%; Seoul, Korea: 48.8%). These observations taken together suggest a compounding effect of urban centralization on scientific research.

**Table 1: A Sample of very low-precision place names.**

| | | |
|---|---|---|
| University, MS, USA | Harvard, MA, USA | Rome, NY, USA |
| Usa, Oita, Japan | Cambridge, WI, USA | Mayo, YT, Canada |
| Institute, WV, USA | Carolina, PR, USA | Sydney, NS, Canada |
| Center, CO, USA | Columbia, NJ, USA | Durham, CT, USA |
| York, NE, USA | Federal, NSW, Australia | King, NC, USA |
| London, KY, USA | Ontario, OR, USA | Melbourne, AR, USA |
| Boston, VA, USA | Denmark, SC, USA | Yale, MI, USA |
| Washington, TX, USA | Poland, ME, USA | Madison, GA, USA |
| Street, Somerset, UK | Hopkins, MN, USA | Wales, WI, USA |
| North, VA, USA | Rochester, MO, USA | Indiana, PA, USA |
| Paris, IL, USA | Florida, NY, USA | Athens, IN, USA |
| Oxford, IA, USA | Mexico, NY, USA | DE, USA; IN, USA |

**Table 2: Ambiguity of the top place names.**

| City | Prec. | Rec. | % of country publications | % of country population |
|---|---|---|---|---|
| London, UK | 91.8 | 91.5 | 29.2 | 13.7 |
| New York, NY | 68.0 | 87.6 | 5.5 | 2.7 |
| Boston, MA | 98.7 | 93.3 | 5.1 | 0.2 |
| Paris, France | 99.0 | 68.7 | 39.5 | 18.5 |
| Tokyo, Japan | 94.7 | 97.4 | 20.7 | 10.2 |
| Beijing, China | 99.4 | 68.7 | 18.2 | 1.6 |
| Seoul, Korea | 97.1 | 99.3 | 48.8 | 19.8 |
| Baltimore, MD | 99.5 | 94.9 | 2.7 | 0.2 |
| Philadelphia, PA | 99.5 | 95.1 | 2.7 | 0.5 |
| Los Angeles, CA | 99.6 | 86.5 | 2.6 | 1.2 |

The top 20 most common countries are shown in Figure 1. Among 37 million affiliations, there are about 11 million in the lower 48 US states and about 3 million in the mainland China during the period of 1988-2016. For each publication, each city is counted once when multiple coauthors are from the same city. As shown in Figure 2, the number of publications has been growing rapidly over time. Note that the y-axis (number of publications) is shown on a log-scale. PubMed started indexing first-author affiliations in 1988 and all authors' affiliations in 2014. Therefore, the availability of geospatial data surges in 1988 and again in 2014. Figure 3 shows the geographical distribution of all papers across all of the United States and its territories. The cities outside the lower 48 US states are geographic outliers and are represented by a small minority of publications, and are therefore excluded from the clustering analysis in this paper.

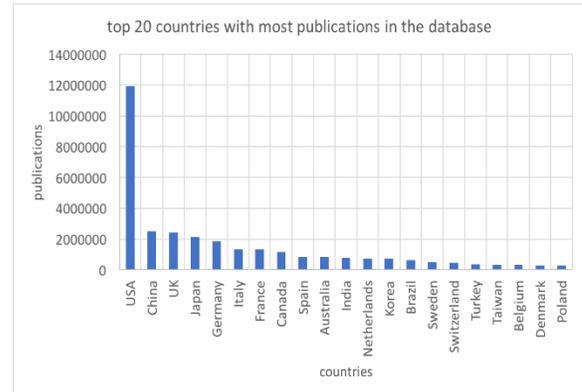

**Fig. 1: The top 20 countries.**

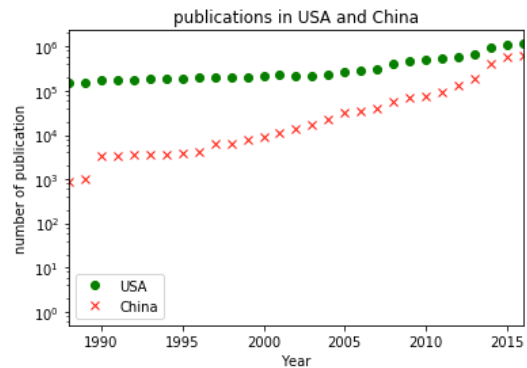

**Fig. 2. Number of papers 1988-2016**.

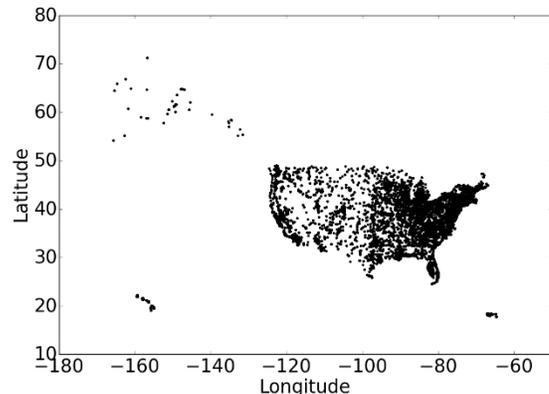

**Fig. 3. Spatial distribution of the USA and its territories.**

## 3 METHODS

### 3.1 Centroids and distances

*3.1.1 Geographical centroid.* For every affiliation in the corpus, the longitude and latitude of its city have been identified and recorded. Given the assumption of Euclidean Geometry, where the longitudes and latitudes are perpendicular to each other and form a plane, for the collections of cities in the Lower 48 states of United States, the central point can be calculated by averaging their latitude and longitude. Localities outside the "Lower 48" (the 48 states other than Alaska and Hawaii) will cause significant influence in calculating the centroid of United States because of the long distance in between. Only the affiliations inside the "Lower 48" are taken into the consideration while calculating the geographical centroid. Those states and territories will be analyzed separately in future work. Furthermore, the method of locating the geographical centroid and calculating the variability in the following section can be applied to not only the "Lower 48", but also USA or other countries, territories and collections of areas.

*3.1.2 Variability calculation.* With the geocode (longitude and latitude) of geographical centroid acquired, the average distance from all cities to the centroid can be calculated. There are different methods to calculate the distance: The Euclidean distance is based on the assumption of flat space; the Great Circle Distance is treating the Earth as a perfect globe and the distance as an arc; the Vincenty distance is based on the assumption of the Earth being an oblate spheroid. In this section, Vincenty distance is selected because of its accuracy to the real situation. Then the distance from each city to the geographical centroid can be calculated by applying the following equation, where $r_i$ is the distance from an individual locality to the centroid, $n$ is the total number of localities in each calculation, $\bar{r}$ is the average distance.

$$\bar{r} = \sqrt{\frac{1}{n} * \sum_i r_i^2}$$

### 3.2 Clustering

*3.2.1 K-Means clustering.* K-Means is a common clustering method in machine learning. In this method, all the data get clustered into k clusters with the k-value predefined by the researcher. First, a group of randomly selected k initial centroids are set, and all the points in the database got clustered to their closest centroid; Then the initial centroids move towards the real centroids of the current clusters, and afterwards, all the data got clustered again with the newly established centroids. The iteration goes until all members are stably clustered in k groups, with the centroids being exactly their centroids. To eliminate the influence of initial centroids, the K-Means clustering will be repeated for several times to check the robustness of the final clustering. In this section, the affiliation cities within the "Lower 48" will be clustered by using K-Means method to demonstrate whether there are some city concentrations or active area in the United States. A change over years will also be provided to manifest the development.

*3.2.2 Determining the number of clusters, k.* The number of clusters, k, has salient significance in the clustering results. The value of k also influence the variability within clusters. When the number of cluster increases, there will be more centroids allocated for all the data points, and the variability, namely the average distance from each point to its nearest centroid will decrease. A threshold on the change of variability can be set before clustering, and the value k is decided when the increase of k does not cause a larger influence than the threshold.

In this section, the effects of the number of clusters will be studied and the average distance within clusters will be estimated. Therefore, the number of clusters and geographical distance can be analyzed for the affiliation localities on the lower 48 United States.

## 4 RESULTS AND DISCUSSION

### 4.1 Geospatial distribution of all publications

Figures 4 and 5 show a density maps of publication rates for the lower 48 US states and mainland China, respectively. The density is represented by the darkness of the color which is drawn on a log-scale shown to the right of each map. From the density map, it can be observed that in the US, there are multiple high-density regions such as along the north-east coast, around Boston-New York-Baltimore, and along the south-west coast, around Los Angeles and San Francisco. In China, the high-density areas are on the east coast, Beijing to the north, and Shanghai and Wuhan to the south. The west of China is sparse.

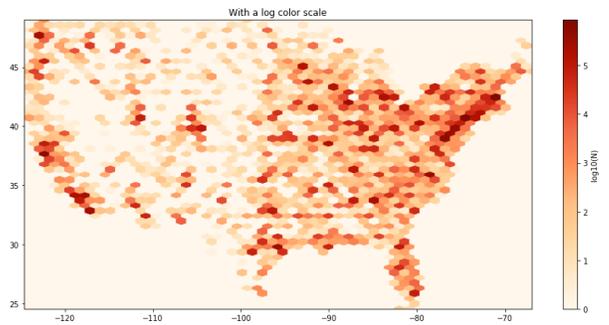

**Fig. 4. Density map of the lower 48 US states 1988-2016.**

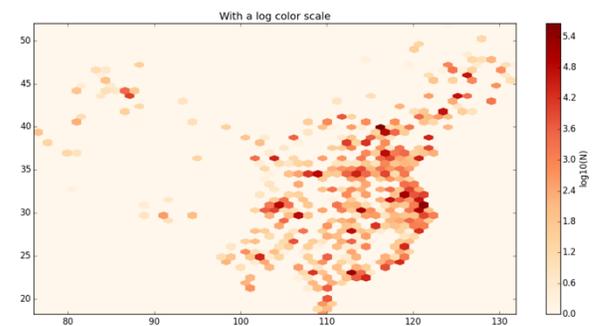

**Fig 5. Density map of mainland China 1988-2016.**

**4.2 Overall centroids and their movements over time**
Figure 6 shows the overall centroids when all papers during 1988-2016 are pooled. The US centroid (-89.2, 38.7) is located in southern Illinois, while the centroid in China (116.2, 34.7) is located in province of Shandong. Figure 7 shows the centroid trends over time. The latitude of the US centroid has been gradually decreasing (moving south) but only about 0.2 degrees (~20 miles) over the past 30 years. Although the longitudes have fluctuated over time by about 0.2 degress, the average longitude has not systematically shifted during this period. In China, the latitude has also decreased but to a much larger extent: 1.7 degrees, most of which occurred during the 1990s, while longitude has fluctuated without a systematic shift in the past 30 years. In other words, the centroids in both countries have moved southward.

effect. However, the average distances represent physical distances on earth (the y-axis represent distances in miles). The red line in the figure 100 miles, and the green line in the figure indicates 50 miles, which are reasonable distances for researchers to travel in a short period of time. In other words, within hub with a radius of 50 or 100 miles, it is reasonable for researchers to form and maintain collaborations with relatively frequent face-to-face meetings. For the USA, it takes only 4 centroids for the average distance to drop below 100 miles, and 6 centroids to drop below 50 miles. The corresponding number of centroids in China are smaller, only 3 and 4 for 100 miles and 50 miles, respectively. Figure 9 shows the corresponding clusters.

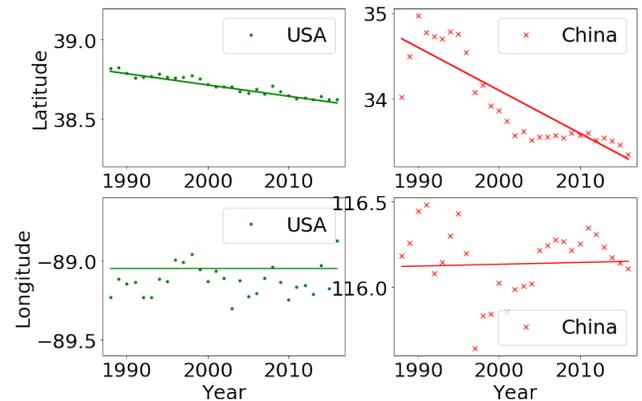

**Fig. 7. Overall centroid movement for the USA (left panels) and China (right panels) over time from 1988 to 2016.**

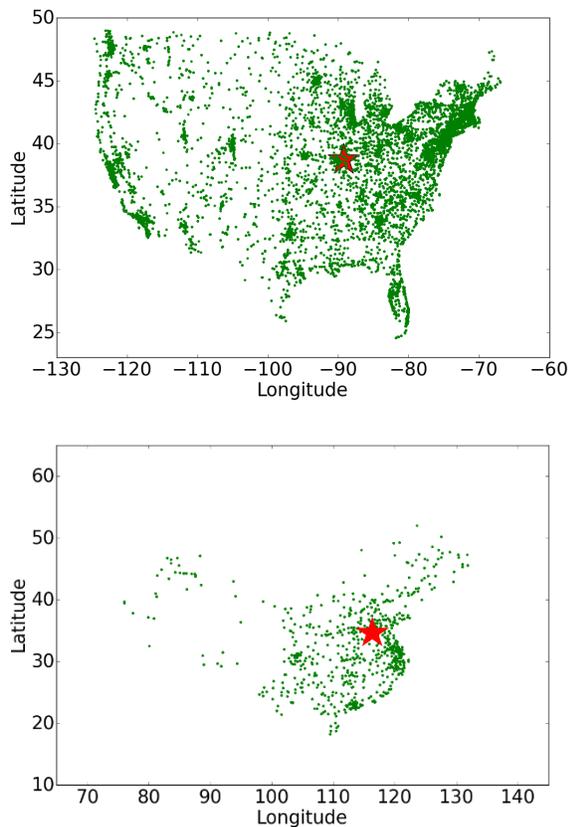

**Fig 6. Overall centroids for the USA (top panel) and China (bottom panel).**

**4.3 Regional clustering**
We now show the results of clustering publications at different levels of granularity using k-means clustering. Here k denotes the number of centroids, each of which capture the notion of a region. Figure 8 shows that the average distance to a regional centroid decreases rapidly (exponentially) as the number of centroids increases, more so for the USA than China because the USA starts out with a much higher average distance for one centroid. This is perhaps not surprising because the data clustering generally has this

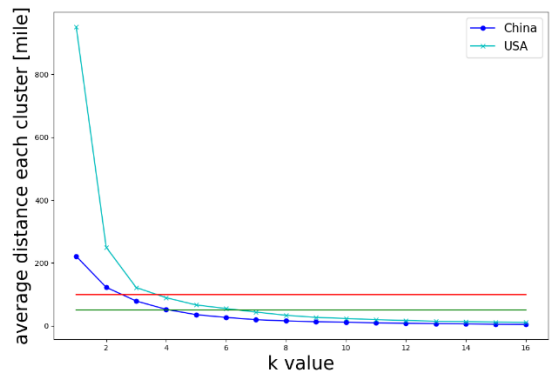

**Fig. 8. Regional clustering: the average distance to the closest centroid decreases rapidly as the number of centroids (k) increases for both the USA and China.**

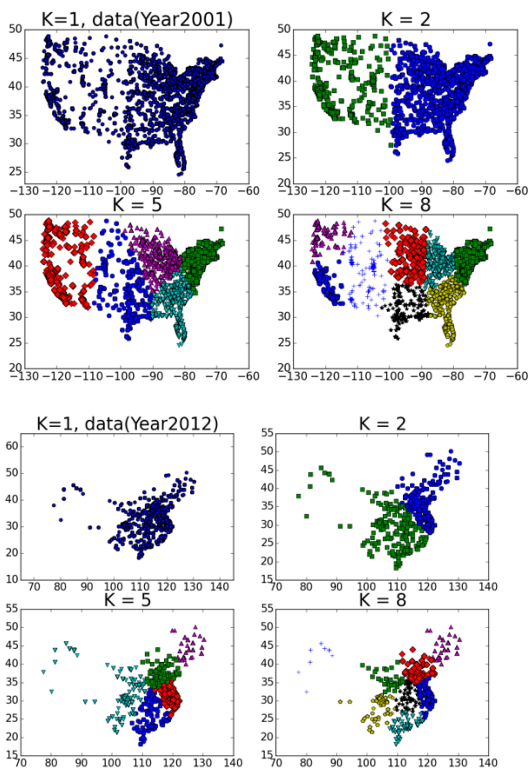

**Fig 9. Regional clusters for different number of centroids (k) for the USA (top panels), and China (bottom panels).**

Another interesting temporal observation is that, although the quantity of publications have increased dramatically, the regional clustering and average distances have remained almost the same. In other words, the quantity within each region has been increasing while the clustering itself has been stable. Figure 10 shows that six regional clusters for the USA in 1988 and 2016, the two temporal extremes in our data, are nearly identical.

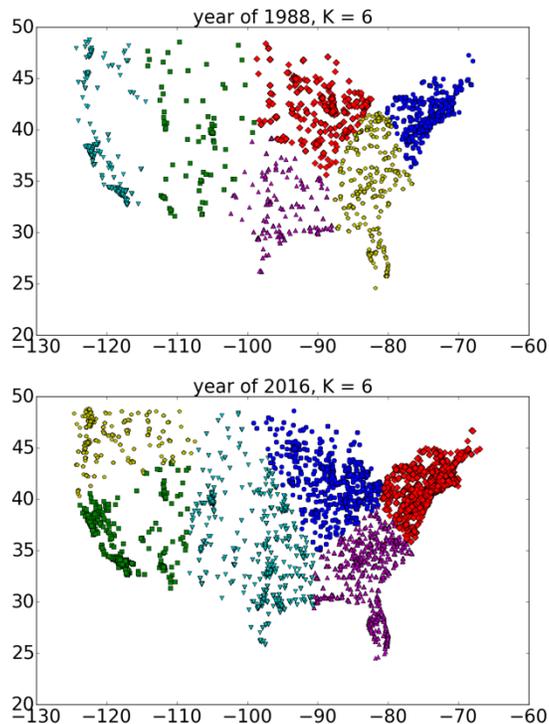

**Fig 10. The regional clustering in 1998 (top panel) and 2016 (bottom panel) in the USA are nearly identical.**

## ACKNOWLEDGMENTS
Research reported in this publication was supported in part by NIH National Institute on Aging P01AG039347. The content is solely the responsibility of the authors and does not necessarily represent the official views of the NIH.